\documentclass[aip,rsi,reprint]{revtex4-1}
\usepackage{graphicx}

\draft 

\begin{document}


\title{The Role of Mode Match in Asymmetric Fiber Cavities} 



\author{A. Bick}
\affiliation{ZOQ (Zentrum f\"{u}r Optische Quantentechnologien)\\Universit\"{a}t Hamburg, Luruper Chaussee 149, 22761 Hamburg, Germany}
\author{C. Staarmann}%
\affiliation{ZOQ (Zentrum f\"{u}r Optische Quantentechnologien)\\Universit\"{a}t Hamburg, Luruper Chaussee 149, 22761 Hamburg, Germany}
\author{P. Christoph}%
\affiliation{ZOQ (Zentrum f\"{u}r Optische Quantentechnologien)\\Universit\"{a}t Hamburg, Luruper Chaussee 149, 22761 Hamburg, Germany}
\author{O. Hellmig}%
\affiliation{ILP (Institut f\"{u}r Laserphysik)\\Universit\"{a}t Hamburg, Luruper Chaussee 149, 22761 Hamburg, Germany}
\author{J. Heinze}
\affiliation{ILP (Institut f\"{u}r Laserphysik)\\Universit\"{a}t Hamburg, Luruper Chaussee 149, 22761 Hamburg, Germany}
\author{K. Sengstock}
\affiliation{ZOQ (Zentrum f\"{u}r Optische Quantentechnologien)\\Universit\"{a}t Hamburg, Luruper Chaussee 149, 22761 Hamburg, Germany}
\affiliation{ILP (Institut f\"{u}r Laserphysik)\\Universit\"{a}t Hamburg, Luruper Chaussee 149, 22761 Hamburg, Germany}
\author{C. Becker}
\email{cbecker@physnet.uni-hamburg.de}
\affiliation{ZOQ (Zentrum f\"{u}r Optische Quantentechnologien)\\Universit\"{a}t Hamburg, Luruper Chaussee 149, 22761 Hamburg, Germany}
\affiliation{ILP (Institut f\"{u}r Laserphysik)\\Universit\"{a}t Hamburg, Luruper Chaussee 149, 22761 Hamburg, Germany}


\date{\today}

\begin{abstract}
We study and realize asymmetric fiber-based cavities with optimized mode match to achieve high reflectivity on resonance. 
This is especially important for mutually coupling two physical systems via light fields, e.g. in quantum hybrid systems.
Our detailed theoretical and experimental analysis reveals that on resonance the interference effect between the directly reflected non-modematched light and the light leaking back out of the cavity can lead to large unexpected losses due to the mode filtering of the incoupling fiber. 
Strong restrictions for the cavity design result out of this effect and we show that planar-concave cavities are clearly best suited.
We validate our analytical model using numerical calculations and demonstrate an experimental realization of an asymmetric fiber Fabry-P\'{e}rot cavity with optimized parameters.
\end{abstract}

\pacs{42.15.Eq, 42.81.Qb, 42.79.Gn, 42.81.-i}

\maketitle 

\section{Introduction}
\begin{figure}[htbp]
  \centering
    \includegraphics{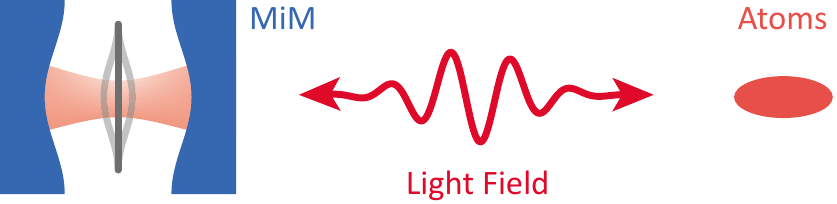}
     \caption{\label{fig:hybrid_sys} \textbf{Sketch of an exemplary quantum hybrid system.} An optomechanical system (here a SiN membrane inside an optical resonator) is coupled to a sample of cold atoms via a light mediated interaction. }
\end{figure}

Fabry-P\'{e}rot cavities have long played an important role in the field of physics. 
They are ideally suited to increase the interaction between light and matter and are an indispensable tool in different fields, e.g. spectroscopy, cavity quantum electrodynamics or more recently in cavity optomechanics\cite{Aspelmeyer2014,Aspelmeyer2014b}.
In these fields, microscopic fiber-based Fabry-P\'{e}rot cavities \cite{Trupke2005,Steinmetz2006a,Hunger2010b,Muller2010a,Greuter2014,Uphoff2014} open interesting new perspectives owing to their unique properties:
One example is their very small mode volume, which allows for stronger coupling to e.g. atomic systems \cite{Gehr2010}.
Another example is the significantly larger  optomechanical coupling strength compared to macroscopic resonators \cite{Flowers-Jacobs2012}.
Furthermore fiber cavities are well-suited for use in sub-Kelvin cryogenic environments \cite{Mamin2001}, since they don't require open apertures to couple light into the cryostat, which minimizes undesired heating effects.\\
For certain applications, e.g. the integral use of such fiber cavities in quantum hybrid systems \cite{Vogell2013,Vogell2014,Joeckel2014}, the most important aspect is a bidirectional light mediated coupling between the two constituents (see Fig.\ \ref{fig:hybrid_sys}).
Here a crucial additional requirement for any cavity enclosing one of the physical systems is a finite reflectivity on resonance.
Symmetric cavities are therefore not suitable for these applications, since they usually exhibit a vanishing reflectivity on resonance.
Obtaining a finite reflectivity is commonly realized by using an asymmetry in the reflectivities of the in- an out-coupling mirrors ($R_{1} \ne R_{2}$).
An important difference between free space cavities and fiber cavities is the lack of possibilities for external mode match, 
which imposes additional severe design restrictions for fiber cavities.
Here the light field emitted by the fiber can not be adapted to the cavity geometry, consequently the cavity geometry has to be adapted to the fiber mode to ensure optimal mode match.
In this case the possible cavity geometry is closely connected to the particular fiber utilized for the fabrication of the cavity, more specifically to the mode field diameter of the guided mode in this fiber.
If the mode match is not perfect, an interference of the intra-cavity field with the promptly reflected non-modematched  field,
leads to a spatial shape of the reflected mode far from a Gaussian. 
In experiments involving free-space cavities this effect is usually less apparent, but leads to a rather unexpected behaviour of the reflected power for asymmetric fiber cavities where the in- and out-coupling fibers act as additional mode filters.
As we show in this paper the key to suppress these unwanted interferences is a close to perfect mode match.
Only a carefully mode matched asymmetric fiber cavity shows a reflection around resonance as expected from the reflectivities of the mirrors.
Minor deviations from perfect mode match already lead to a drastic decrease of the reflected power on resonance.
This is in harsh contrast to free-space cavities, where an imperfect mode match results in a higher reflectivity on resonance.

In the following we study theoretically and experimentally the optimal geometry for asymmetric fiber-based cavities. 
We show that planar concave cavities are best suited and present a realization of a home-built fiber cavity with optimized parameters.
In particular the radius of curvature of the curved fiber tip as well as the range of usable cavity lengths has been adapted to the specific fiber to maximize the mode match, leading to the desired reflectivity on resonance.
Our studies pave the way for the implementation of atom-membrane hybrid quantum systems in a low milli-Kelvin cryogenic environment,
which shall allow sympathetic cooling of the membrane to its quantum mechanical ground state by laser cooling the atoms \cite{Vogell2013,Vogell2014}.

The  paper is organized as follows:
After a short reminder on basic properties of Fabry-P\'{e}rot cavities, we introduce an analytical model to calculate reflection and transmission properties of asymmetric fiber cavities and discuss the role of mode matching. 
We then turn to the determination of optimal cavity parameters for a given fiber and design wavelength and show how slight deviations from these optimal values already produce large
variations of the reflected power with respect to the expected values.
Finally we present an experimental characterization of an optimized asymmetric fiber cavity that we produce in-house.

\section{Properties of Fabry-P\'{e}rot Cavities}
\label{sec:fpcavprop}

\begin{figure}[htbp]
  \centering
    \includegraphics{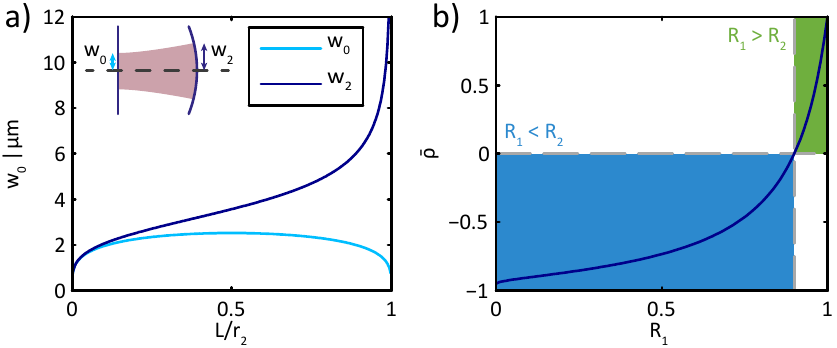}
     \caption{\label{fig:fc_w0w1} \textbf{Properties of planar concave cavities.} 
     \textbf{a)} Spot sizes $w_0$ and $w_2$ on the mirrors of a planar concave cavity (see inset) with a radius of curvature of $r_2 = 50 \, \mu \mathrm{m}$. For planar concave cavities the waist $w_0$ is located on the planar mirror.
     \textbf{b)}~Field reflectivity on resonance $\bar \rho$ plotted for a fixed reflectivity of the outcoupling mirror of $R_2 = 0.9$.}
\end{figure}

The radius of curvature and the distance between the mirrors of a Fabry-P\'{e}rot cavity define the spatial cavity mode characterized by  a waist $w_0$ and spot sizes on the incoupling $w_1$ and outcoupling mirror $w_2$ as shown in Fig \ref{fig:fc_w0w1} a. 
If the cavity is resonant with the light field ($\phi = n \cdot 2\pi,\, n \in \mathbf{N}$), a dip in the reflected intensity
\begin{equation}
\frac{I_\mathrm{ref}}{I_\mathrm{in}} = \left| \frac{E_\mathrm{ref}}{E_\mathrm{in}} \right|^2 = \left| \frac{\sqrt{R_1} - e^{2 i \phi} \sqrt{R_2}}{1 - e^{2 i \phi}\sqrt{R_1 R_2}} \right|^2\label{eq:Iref}.
\end{equation}
occurs. 
$E_i$ denote the electric field values of the involved fields and $\phi$ is the phase picked up by the light field during one cavity round-trip.
The depth of this dip is characterized by the \textbf{field reflectivity on resonance $\bar \rho$}:
\begin{equation}
\bar\rho = \frac{\sqrt{R_1} - \sqrt{R_2}}{1 - \sqrt{R_1 R_2}},
\label{eq:rhobar}
\end{equation}
Here $R_1$ and $R_2$ is the power reflectivity of the in- and outcoupling mirror respectively (compare Fig.\ \ref{fig:fc_w0w1}b).
We will employ $\bar{\rho}$ as a central quantity throughout this paper.\\
For a symmetric cavity ($R_1 = R_2$) $\bar \rho$ vanishes, while  $\bar \rho$ turns positive if the incoupling mirror has a higher reflectivity ($R_1 > R_2$).
Note, that experiments typically measure the power reflectivity $\rho = \bar \rho^2$ (see Eq.\ \ref{eq:Iref}).

The corresponding transmission through the cavity is described by the field transmittance on resonance
\begin{equation}
\bar \tau = \frac{\sqrt{(R_1 - 1)(R_2 - 1)}}{1 - \sqrt{R_1 R_2}}\label{eq:tau}
\end{equation}
with the transmitted power $\tau = \bar \tau^2$.
\section{Asymmetric Fiber Cavities} 
\label{sec:asymfc}
It is well-known, that to couple light efficiently into an optical fiber, the incoming beam and the fiber mode have to be mode matched. 
For a fiber-based cavity this means that the reflected and transmitted fields of the cavity ideally have to match the respective fiber modes to yield the expected power reflectivity and transmission measured at the end of the in- and out-coupling fibers.\\
\begin{figure}[t]
  \centering
    \includegraphics{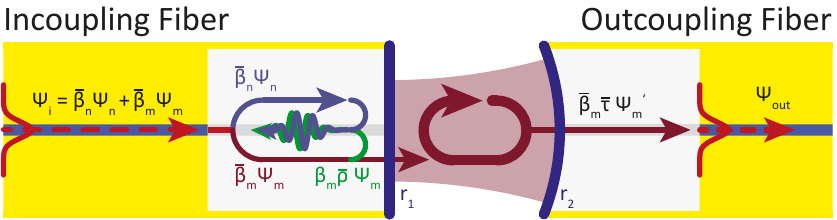}
     \caption{ \label{fig:fcmatch} \textbf{Sketch of the fields coupled into a fiber Fabry-P\'{e}rot cavity.} 
     The gray area indicates the virtual zone between the end of the fiber and the coating. 
     The input field $\psi_\mathrm{i}$ is split into a mode-matched part $\psi_\mathrm{m}$ and a non-mode-matched part $\psi_\mathrm{n}$. 
     While the first enters the cavity and is subsequently reflected with the field reflectivity on resonance $\bar \rho$, the latter is non-resonant and directly reflected. 
     These two fields interfere. 
     The field transmitted through the cavity $\psi_\mathrm{m}^\prime$ has to be matched with the mode of the outcoupling fiber $\psi_\mathrm{out}$.}
\end{figure}

Fig.\ \ref{fig:fcmatch} schematically shows  a fiber Fabry-P\'{e}rot cavity with imperfect mode match.
The input field $\psi_\mathrm{i}$, guided in the incoupling fiber, can be decomposed into a part matched with the cavity ground mode $\psi_\mathrm{m}$ and a non-mode-matched part $\psi_\mathrm{n}$:
\begin{equation}
	\psi_\mathrm{i} = \bar \beta_\mathrm{n} \psi_\mathrm{n} + \bar \beta_\mathrm{m} \psi_\mathrm{m}. \label{eq:psii}
\end{equation}
This is valid as the higher spatial modes are in general non-resonant with the cavity already for a small finesse. 
They are reflected off the cavity with the off-resonant reflectivity, which we assume at this point to be equal to one.\\
The parameters $\bar \beta_i$, introduced in the field decomposition (see Eq. \ref{eq:psii}), are complex numbers, which account for the mode match. 
The fields $\psi_\mathrm{i}$, $\psi_\mathrm{n}$, and $\psi_\mathrm{m}$ are normalized
\begin{equation}
	\int \left| \psi_\mathrm{i} \right|^2 \mathrm{d}A = \left| \bar \beta_\mathrm{n} \right|^2 \int \left| \psi_\mathrm{n} 	\right|^2 \mathrm{d}A + \left| \bar \beta_\mathrm{m} \right|^2 \int \left| \psi_\mathrm{m} \right|^2 \mathrm{d}A = 1,
\end{equation}
yielding $\beta = \left| \bar \beta_\mathrm{m} \right|^2 = \left| \bar \beta_\mathrm{n} \right|^2 - 1$.
 $\beta = 1$ thus corresponds to perfect mode match.\\

In the following we study the following parameters in detail to understand the reflection and transmission characteristics of asymmetric fiber-based cavities:
\begin{enumerate}
	\setlength{\itemsep}{0pt}
	\item \textbf{The power reflectivity} given by the sum of the directly reflected non-mode-matched part and the light reflected by the cavity. See Section \ref{sec:powerref}.
	\item \textbf{The reflection mode match} is crucial to understand how the interference of the non-mode-matched field and the field reflected by the cavity, when filtered by the fiber mode influences the reflected power measured at the end of the incoupling fiber. See Section \ref{sec:refmm}.
	\item \textbf{The power transmittance} which we calculate from the reflected power if losses are ignored and which is less influenced by mode mismatch. See Section \ref{sec:ptrans}.
	\item \textbf{The transmission mode match} finally governs the power measured at the end of the out-coupling fiber. See Section \ref{sec:modetrans}.
\end{enumerate}

\subsection{Power Reflectivity}
\label{sec:powerref}

In this subsection we derive and explain an expression for the reflected power that explicitly depends on the mode match of incoupling fiber and cavity mode.
Together with the results on the reflection mode match itself, derived in the next subsection \ref{sec:refmm}, it will become clear why an imperfect mode match can completely hinder the envisioned functionality of an asymmetric fiber cavity.\\
The power of a field $\psi$ can be calculated via
\begin{equation}
	P = \int \left| \psi \right|^2 \mathrm{d}A.
\end{equation}
The mode-matched part of the light (the field that is interacting with the cavity) is reflected with the field reflectivity on resonance $\bar \rho$ as described by Eq.\ \ref{eq:rhobar}. 
This part interferes with the non mode-matched part that is promptly reflected from the cavity as sketched in Fig.\ \ref{fig:fcmatch}. 
The reflected field $\psi_\mathrm{r}$ is therefore:
\begin{equation}
	\psi_\mathrm{r} =  \bar \beta_\mathrm{n} \psi_\mathrm{n}  + \bar\rho \bar \beta_\mathrm{m} \psi_\mathrm{m}.  \label{eq:psir}
\end{equation}
The reflected power $P_\mathrm{ref}$ can therefore be calculated as
\begin{eqnarray}
P_\mathrm{ref} &=& \int \left| \psi_\mathrm{r} \right|^2 \mathrm{d}A \\
&=& \int \left|\bar \beta_\mathrm{n}\right|^2 \left| \psi_\mathrm{n} \right|^2 +  \left|\bar \rho \bar \beta_\mathrm{m}\right|^2 \left| \psi_\mathrm{m} \right|^2 \\
&+& 2 \bar\rho \cdot \Re[ \bar \beta_\mathrm{n} \bar \beta_\mathrm{m}^*  \psi_\mathrm{m} \psi_\mathrm{n}] \mathrm{d}A
\label{eq:spatialmode}\\
&=& 1 - \beta \left( 1 - \rho \right). \label{eq:pref}
\end{eqnarray}
The matched and not-matched functions are orthogonal, therefore also the integral of the real part is zero. 
$P_\mathrm{ref}$ is as expected from an asymmetric cavity \cite{Trupke2005}.
A closer look at the reflected spatial intensity distribution (Eq. \ref{eq:spatialmode}) shows that the mixed term,
which drops out during the integration, has an effect on the spatial mode of the reflected field. 
Note that the field reflectivity on resonance $\bar \rho$ changes sign when the role of the mirrors $R_1$ and $R_2$ is reversed. 
For a regular free-space cavity this is not very important, since the total reflected power $P_\mathrm{ref} \propto \rho  = \bar \rho ^2$ is typically measured with a photo diode and thus not affected by this sign change.
It implies however dramatic modifications for the spatial mode of the reflected field. 
In the case of a fiber cavity, this reflected mode $\psi_\mathrm{r}$ has to match the fiber mode to be guided through and finally measured at the end of the fiber.
\subsection{Reflection Mode Match}
\label{sec:refmm}
\begin{figure}[htbp]
  \centering
    \includegraphics{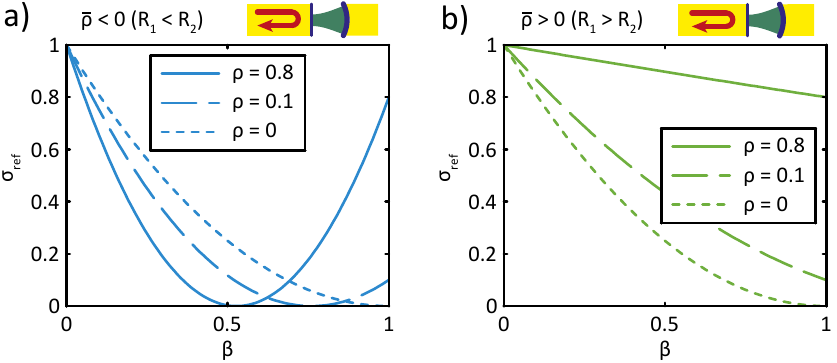}
     \caption{\label{fig:fc_sigma} \textbf{Observed reflection signal $\sigma_\mathrm{ref}$ for an asymmetric cavity as a function of mode match $\beta$.} 
     \textbf{a)} The incoupling mirror has  smaller reflectivity. 
     \textbf{b)} The outcoupling mirror has  smaller reflectivity. 
     If $\rho = 0$, $R_1 = R_2$ and both lines overlap.}
\end{figure}
The power coupling efficiency of the reflected field $\eta_\mathrm{ref}$ is given by the overlap integral of the reflected field $\psi_\mathrm{r}$ (Eq. \ref{eq:psir}) and the fiber mode $\psi_\mathrm{i}$ (Eq.\ \ref{eq:psii}):
\begin{equation}
\eta_\mathrm{ref} = \left| \frac{1}{\sqrt{P_\mathrm{ref}}}  \int \psi_\mathrm{r} \psi_\mathrm{i}^* \mathrm{d}A \right|^2 
= \frac{1}{P_\mathrm{ref}}   \left| 1- \beta + \bar\rho \beta \right|^2.
\end{equation}

Multiplying the reflected power $P_\mathrm{ref}$ with the mode match factor $\eta_\mathrm{ref}$ yields the reflection signal $\sigma_\mathrm{ref}$ that can be measured at the end of the incoupling fiber:
\begin{equation}
	\sigma_\mathrm{ref} = \eta_\mathrm{ref} P_\mathrm{ref} =  \left| 1-\beta + \bar \rho \beta  \right|^2 \label{eq:sigmaref}.
\end{equation}
This important result is shown in Fig.\ \ref{fig:fc_sigma} .
Depending on the sign of $\bar \rho$ the observed behavior is strikingly different. 
For an asymmetric cavity, coupled from the lower reflective side ($\bar{\rho }<0$, see Fig.\ \ref{fig:fc_sigma}a), the interference effect can change the spatial mode of the field reflected by the cavity in such a way that the signal observed at the end of the fiber, $\sigma_\mathrm{ref}$ is zero, although the reflectivity on resonance defined by the mirrors $\rho$ has a finite value. 
The measured reflected power is lower than in the free space case -- the asymmetric cavity looks like a symmetric cavity. 
Only for perfect mode match ($\beta = 1$), the expected reflectivity on resonance $\rho$, defined by the asymmetry of the mirrors can be observed. 
Note that if an asymmetric fiber cavity is operated from the higher reflective side \cite{Trupke2005,Greuter2014} ($\bar{\rho} >0$, see Fig.\ \ref{fig:fc_sigma}b) the effect described above becomes much less important and the reflectivity on resonance scales almost linearly with the mode match.
Here $1 \ge \sigma_{\mathrm{ref}} \ge \rho$,
meaning in particular that the cavity will never look symmetric and will never exhibit vanishing reflectivity on resonance.\\
\begin{figure}[t]
  \centering
    \includegraphics{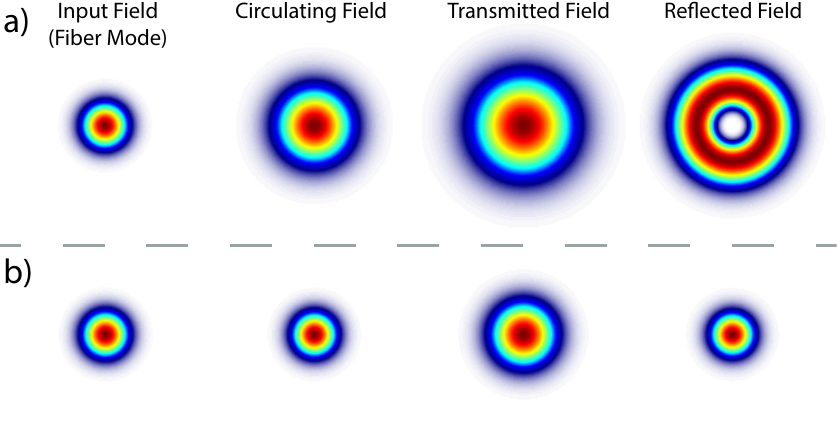}
     \caption{\label{fig:cavityfields} \textbf{Normalized fields of an asymmetric planar concave cavity}, calculated with OSCAR (see App. \ref{sec:oscar}).
		\textbf{a)} Asymmetric cavity with bad mode match ($r_2 = 150\,\mu \mathrm{m}$, $L = 30\,\mu \mathrm{m}$).
      \textbf{b)} Asymmetric cavity with stable mode match ($r_2 = 50\,\mu \mathrm{m}$, $L = 25\,\mu \mathrm{m}$). 
      The input, circulating, and the reflected mode are calculated at the position of the incoupling mirror, whereas the transmitted mode is located at a position directly behind the outcoupling mirror. 
      The incoupling field has a waist of $w =2.5\,\mu \mathrm{m}$.}
\end{figure}
The importance of a close to perfect mode match is  further emphasized in Fig. \ref {fig:cavityfields},
where the involved fields are exemplary shown for an asymmetric fiber cavity with $\bar{\rho} <0$.
A comparison of the input field with the reflected field illustrates the necessity for optimal mode match for asymmetric fiber cavities:
In the case of bad mode match, the overlap of input field and reflected field can even vanish (Fig. \ref {fig:cavityfields}a) leading to vanishing reflectivity on resonance.
\subsection{Power Transmittance}
\label{sec:ptrans}
The transmitted power through the cavity is given by the product of the incoupling mode match $\beta$ and the power transmittance $\tau$ of the asymmetric resonator:
\begin{equation}
P_\mathrm{trans} = \beta \tau.\label{eq:}
\end{equation}
\subsection{Transmission Mode Match}
\label{sec:modetrans}
To obtain the measurable signal at the \textbf{end of the out-coupling fiber} the transmitted power has to be multiplied by the transmission mode match.
The transmitted field is given by $\bar \beta_\mathrm{m} \bar\tau\cdot \psi_\mathrm{m}^\prime$. 
Since no interference is present, one obtaines a more intuitive result:
\begin{eqnarray}
\eta_\mathrm{trans} &=& \left| \frac{1}{\sqrt{P_\mathrm{trans}}}  \int \left( \bar \tau  \bar \beta_\mathrm{m} \cdot \psi_\mathrm{m}^\prime \right) \cdot \psi_\mathrm{out}^* \mathrm{d}A \right|^2\\
&=& \left| \int \psi_\mathrm{m}^\prime \psi_\mathrm{out}^* \mathrm{d}A \right|^2.
\end{eqnarray}
The total observed transmission signal is therefore:
\begin{equation}
\sigma_\mathrm{trans} = \eta_\mathrm{trans}P_\mathrm{trans} = \eta_\mathrm{trans} \beta \tau \label{eq:sigmatrans}.
\end{equation}
The Equations \ref{eq:sigmaref} and \ref{eq:sigmatrans} are expressions for the transmission and reflection of an asymmetric fiber Fabry-P\'{e}rot cavity. 
They depend on the mode match $\beta$ of the cavity mode with the mode emitted by the fiber and the overlap of the cavity mode and the mode of the outcoupling fiber $\eta_\mathrm{trans}$. 
From an experimental point of view it is important to explicitly evaluate the dependence of the mode match on relevant experimental parameters such as radius of curvature of the mirrors, cavity length, radial displacement, etc. 
This is discussed in the App.~\ref{sec:gau_beam_overlap}.
\section{Optimal Cavity Parameters}
\begin{figure}[htbp]
  \centering
    \includegraphics{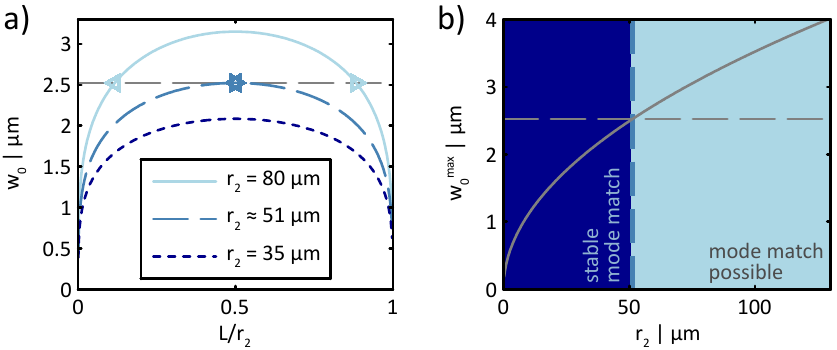}
     \caption{\label{fig:fc_waist}\textbf{Mode match for planar concave cavities. }
     \textbf{a)} Waist of a planar concave cavity for different radii of curvature of the curved mirror $r_2$ as a function of the normalized cavity length  $L/r_2$ . 
     The maximal waist of the cavity is always located at $L = r_2/2$. 
     The dashed line indicates the waist of light at 780 nm emitted from a single mode  optical fiber with a numerical aperture of 0.12 .
     $\triangleleft$/$\triangleright$ indicate the first/second mode match point.
     \textbf{b)} Maximal waist at $L = r_2 / 2$ of a planar concave Fabry-P\'{e}rot cavity as a function of  the radius of curvature of the second mirror $r_2$. 
     The horizontal dashed line shows the same as in a).}
\end{figure}
The necessity of optimal mode match stressed in the preceding chapter results in strong restrictions for the cavity geometry.
In a concave-concave (CC) cavity, $\beta = 1$ is not possible due to the lensing effect of the curved incoupling fiber \cite{Hunger2010b}. 
It results in a stronger divergence, hence moving the virtual waist $w_0$ of the emitted beam into the fiber. 
The fiber and the cavity mode may have the same spot size on the incoupling mirror $w_1$ but the different sign of the wave front curvature $R$ always results in $\beta < 1$. 
In a planar concave (PC) fiber cavity $R = \infty$ is always valid at the planar incoupling mirror. 
Therefore $\beta = 1$ can be realized if the spot emitted by the fiber $w_\mathrm{f}$ has the same size as the cavity mode $w_1 = w_0$ on the incoupling mirror. 
Note that CC cavities may still be the best choice for applications where exceedingly small mode volumes are required, e. g. experiments with single atoms strongly interacting with a cavity mode.
In that case the optimal parameter range has to be carefully evaluated balancing mode match and coupling strength. 
For our envisioned application of a MiM system coupled to cold atoms, an ever smaller mode volume is of little interest but we require the highest possible reflectivity on resonance, 
or in other words the best possible mode match.
Therefore we will concentrate on PC geometries first and briefly discuss the correspondning CC results subsequently.\\

The waist $w_0$ in a PC cavity depends on the length $L$ and the radius $r_2$. 
Furthermore, $w_0$ has a maximum at $L = r_2/2$ (see Fig.\ \ref{fig:fc_waist}a). 
Only if the waist of the fiber is smaller or equal than this maximum, perfect mode match, i.e. $\beta = 1$ is possible. 
The intersection points where $w_\mathrm{f} = w_0$ will be referred to as \textbf{mode match points}. 
The first mode match point is close to $L = 0$, the second mode match point close to $L = r_2$. 
If $w_0\left(L = r_2/2\right) = w_\mathrm{f}$, both mode match points fall together and the cavity has a \textbf{stable mode match}. 
In this geometry a length change only weakly affects the cavity waist as $w_0$ has a maximum. 
In the following we exemplarily calculate cavity parameters for a polarization maintaining fiber with  $N\!A=0.12$ on the incoupling side and a single mode fiber with $N\!A=0.13$ at the outcoupling side operated at $780\,\mathrm{nm}$ as used in our experimental realization of asymmetric fiber cavities (see Sec. \ref{sec:fc_expres}).
For such an incoupling fiber the above described stable mode match corresponds to a radius of curvature of $r_2 = 51.28\,\mu \mathrm{m} = r_\mathrm{s}$ (see Fig. \ref{fig:fc_waist}b).

\begin{figure}[t]
  \centering
    \includegraphics{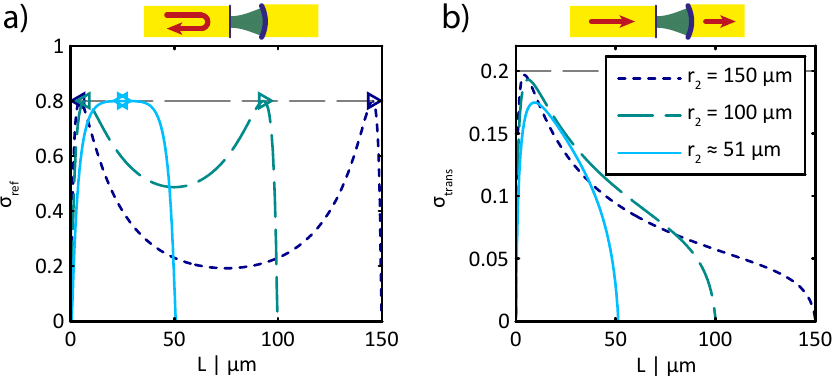}
     \caption{\label{fig:fc_measured}\textbf{Calculated reflection and transmission signals $\sigma$ of an asymmetric PC fiber cavity.} 
     \textbf{a)} Measurable reflected power at the end of the in-coupling fiber $\sigma_\mathrm{ref}$. 
     \textbf{b)} Measurable transmission signal at the end of the out-coupling fiber $\sigma_\mathrm{trans}$. 
     The dashed gray line indicates the power reflectivity on resonance $\rho = 0.8$ in \textbf{a)} and $\ 1 - \rho$ respectively in \textbf{b)}. 
     $\triangleleft$/$\triangleright$ indicate the first/second mode match point. 
     Both fibers have $N\!A = 0.12$ and $\lambda = 780\,\mathrm{nm}$ is used.}
\end{figure}

\subsection{Reflection Signal}
To fulfill the requirement of highest possible reflectivity on resonance an asymmetric fiber cavity can \textbf{only} be operated at the mode match points. 
This is illustrated in Fig.\ \ref{fig:fc_measured}a. 
The behavior of $\sigma_\mathrm{ref}$ is completely dominated by the interference effect described in Sec. \ref{sec:asymfc}. 
Only at the mode match point, the light reflected by the cavity is guided in the incoupling fiber and the reflectivity on resonance $\rho$, given by the reflectivities of the mirrors, is achieved. 
For a cavity mode matched at an unstable mode match point, the length has to be controlled very precisely as the peak of $\sigma_\mathrm{ref}$ as a function of $L$ is very narrow. 
Depending on the radius $r_2$, the observed signal $\sigma_\mathrm{ref}$ can drop to zero in between the mode match points.\\

Note that the results presented here do not include intra-cavity losses $l$. 
For long cavities, for example at the second mode match point for a cavity with unstable mode match, the spot size of the cavity mode on the outcoupling mirror $w_2$ can get very large. 
This can lead to non-negligible clipping losses due to the finite size of the micro-fabricated curved mirror at the fiber tip, ultimately limited by the fiber diameter. 
These losses can be approximately modeled as an effective lower reflectivity of the outcoupling mirror in the form $R_2^\prime = R_2(1 - l)$.
At a given cavity length $L$ where $R_2^\prime = R_1$ the asymmetric cavity will thus turn into a symmetric cavity with vanishing reflectivity on resonance.
Because of this effect the range of useful cavity lengths suitable for experiments requiring high reflectivity on resonance has to be carefully evaluated and depends on different parameters,
most importantly the cavity mode and the size and degree of perfection of the curved mirror on the outcoupling fiber tip.
We will come back to this point in Sec. \ref{sec:fc_expres} where we present our experimental results.

\subsection{Transmission Signal}
The transmission of the unstable and stable mode matched cavity exhibit a similar behavior since no interference effect is present (see Fig.\ \ref{fig:fc_measured}b). 
With increasing length, the spot size on the outcoupling mirror $w_2$ increases, hence the transmission mode match decreases.
In addition the aforementioned lensing effect of the curved fiber surface also negatively affects the transmission mode match. \\
Consequently the resulting very low transmission for large cavity lengths $L$ further limits the usability of the second mode match point.
\subsection{Maximal Length}
As a result of the above considerations the stable matched cavity is the longest possible mode-matched PC cavity usable for experiments.
Although a perfect mode match is possible at the first mode match point, the corresponding cavity length $L$ becomes very small as $r_2$ increases (see Fig.\ \ref{fig:fc_measured}a).
The maximal length depends on the fiber and wavelength used.
Unfortunately, the change in position of the first mode match point with respect to the radius of curvature of the curved mirror $r_2$ is very large.
If $r_2$ is $10\,\mu \mathrm{m}$ larger than the optimal radius $r_\mathrm{s}$,
the first mode match point moves from $L \approx 25\,\mu \mathrm{m}$ to $L \approx 14\,\mu \mathrm{m}$ for a fiber with $N\!A = 0.12$.\\
If a long cavity is not required and the cavity length can be precisely controlled, it can be beneficial to operate a cavity at the first unstable mode match point.
In such a configuration the transmission of the cavity can be substantially better than for a stable mode matched cavity (see Fig.\ \ref{fig:fc_measured}a and b, here the length corresponding to maximum transmission and the first mode match point almost coincide for $r_2 = 150 \, \mu \mathrm{m}$).
To give an example, a transmission mode match close to unity is possible in a geometry where $r_2 = 200\,\mu \mathrm{m}$ and $L = 3.5\,\mu \mathrm{m}$.

\subsection{Planar Concave \& Concave Concave Cavities}
In CC geometries with $r_1 = r_2 = r$, the mode match $\beta^\mathrm{CC}$ is always smaller than one due to the lensing effect of the incoupling fiber. 
Furthermore, the mode match $\beta$ for a CC cavity exhibits only one maximum in contrast to a PC cavity. 
The spot size on the incoupling mirror $w_1$ increases monotonically with the cavity length $L$ whereas the waist $w_0$ has a maximum at $L=r_2/2$ (see Fig. \ref{fig:fc_w0w1}).
With increasing radius $r_1$, the maximum of $\beta^\mathrm{CC}$ increases as the lensing effect decreases.
However  the longest realization of a CC cavity where $\beta^\mathrm{CC}$ exhibits a maximum is given by  $L \approx 13\,\mu \mathrm{m}$ and $r \approx 50\,\mu \mathrm{m}$ for a fiber with $N\!A = 0.12$ fiber. 
Note, that in this configuration the maximum of $\beta^\mathrm{CC} = 0.89$ is still severely limited by the lensing effect and the length is only half that of the stable matched PC geometry. 
For large radius $r$, CC and PC cavities show an identical behavior but at $L^\mathrm{CC} = 2 L^\mathrm{PC}$.\\

The findings can be summarized as: 
(1) A stable matched PC cavity is the longest practicably usable asymmetric fiber cavity and has a length $L=r_2/2$.
(2) If high transmission or high incoupling efficiency from the higher reflective side are important it is beneficial to use the first stable mode match point and $r_2 > 200\,\mu \mathrm{m}$. (3) A CC geometry increases the cavity length by a factor of two with little negative effect for the incoupling efficiency if $r_2 > 200\,\mu \mathrm{m}$, while still being significantly shorter than the PC cavity with stable mode match.

\section{Experimental Realization}
\label{sec:fc_expres}
In the following section experimental results of the realization of an optimized asymmetric fiber Fabry-P\'{e}rot cavity are presented.
The cavity is made out of a planar polarization maintaining fiber with $N\!A = 0.12$\cite{Note1} on the incoupling side
and a curved single mode fiber with $N\!A = 0.13$\cite{Note2} at the outcoupling side.
We produce the concave features at the curved fiber tip in-house by CO$_2$ laser machining.
Subsequently we coat the fiber tips with soft coatings to adjust the reflectivity on resonance of the envisioned asymmetric cavity.

\begin{figure}[t]
  \centering
    \includegraphics{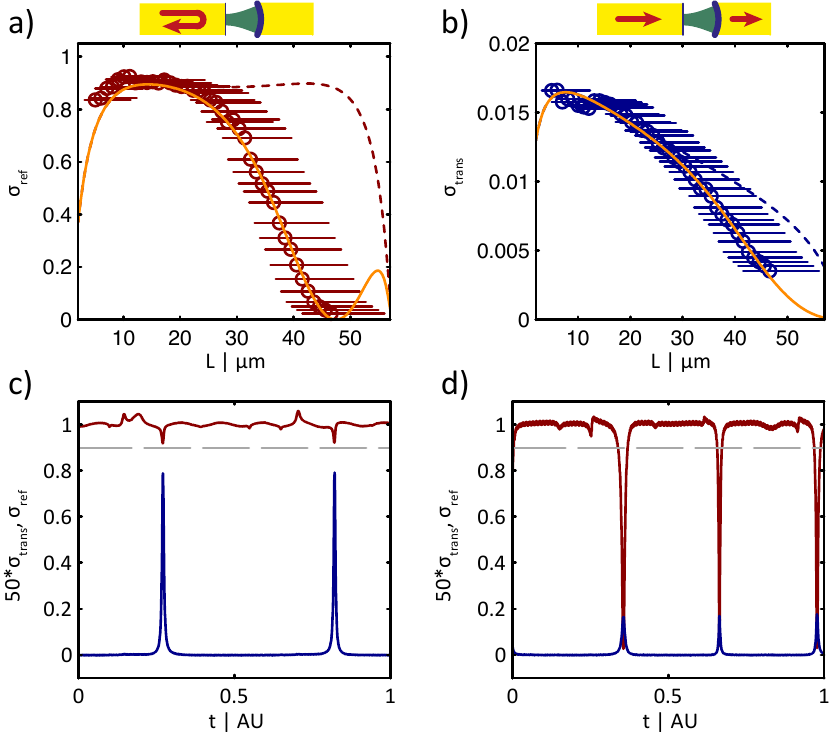}
     \caption{\label{fig:fc_sym_asym} \textbf{Experimental realization of an optimized asymmetric PC fiber Fabry-P\'{e}rot cavity} with $R_1~=~0.928$ and $R_2~=~0.9982$, corresponding to a Finesse $\mathcal{F}~\approx 80$. 
     \textbf{a)} Reflection signal $\sigma_\mathrm{ref}$. 
     The orange line indicates the fitted reflection signal if clipping losses on the curved mirror with $r^\mathrm{ref}_2 = 58\,\mu \mathrm{m}$ and an effective diameter $D_\mathrm{ref} = 12.8\,\mu \mathrm{m}$ are included (see text).
     The red dashed line is the theoretical model without losses for $r^\mathrm{ref}_2 = 58\,\mu \mathrm{m}$. 
     Data points are the extracted reflection values from a scan across the cavity resonance. 
     The error bars indicate the error of the micrometer drive calibration and the cavity length underestimation caused by a small inclination of $2^{\circ}$ between optical axis of the imaging system and the cavity gap.
     \textbf{b)} Corresponding measured transmission signal $\sigma_\mathrm{trans}$ together with the fitted analytical model with losses (orange) for $r^\mathrm{trans}_2 = 60\,\mu \mathrm{m}$ and $D_\mathrm{trans} = 13.0\,\mu \mathrm{m}$ and without clipping losses (dashed blue). 
     The modeled transmission is scaled by a factor of 0.18 (see text).
     \textbf{c)} Scans of the cavity length for $L = 8 {+ 6.5 \atop -2} \, \mu \mathrm{m}$.
     Shown in blue/red is the transmission/reflection signal. 
     The transmission signal is scaled by a factor of 50. 
     The dashed gray line indicates the expected reflectivity on resonance of $\rho \approx 0.9$.
     \textbf{d)} As left but for $L = 45 {+ 9.5 \atop -5} \, \mu \mathrm{m}$.}
\end{figure}

The corresponding measured transmission and reflection signals of this optimized cavity are shown in Fig.\ \ref{fig:fc_sym_asym}. 
Each data point corresponds to a scan of the cavity length over at least two cavity resonances.
From these traces, of which two are exemplarily shown in Fig.\ \ref{fig:fc_sym_asym}c and d, the reflection and transmission on resonance ($\sigma_\mathrm{ref}$ and $\sigma_\mathrm{trans}$), and the finesse are extracted. 
For each length, we optimize the transversal cavity alignment prior to the measurements.\\

We find very good quantitative agreement with the theoretical model for short and intermediate cavity lengths ($L < 25\,\mu \mathrm{m}$). 
In this regime our home built asymmetric fiber cavity can be used as intended and exhibits the expected reflectivity on resonance $\rho$.\\
We determine the effective radius of the curved mirror $r^\mathrm{eff}_2 = \left( r^\mathrm{ref}_2 + r^\mathrm{trans}_2 \right)/ 2 =  59\,\mu \mathrm{m}$ and its effective diameter $D_\mathrm{eff} = \left( D_\mathrm{ref} + D_\mathrm{trans} \right)/2 =  12.9\,\mu \mathrm{m}$ by fitting our model (Eq.\ \ref{eq:sigmaref} and \ref{eq:sigmatrans}) to the $\sigma_\mathrm{ref}$ and $\sigma_\mathrm{trans}$ data\cite{Note3}.\\
We independently determine the reflectivity of $R_2$ by measuring the finesse of a CC cavity with two identically coated fibers yielding  $\mathcal{F}=1780$ and $R_2 = 0.9982$ respectively.
The reflectivity of $R_1 = 0.928$ is derived from the median of the measured finesse for $L < 20\,\mu \mathrm{m}$, $\mathcal{F} \approx 80$. 
The length of the cavity is determined by calibrating the differential drive using a microscope camera with a measured resolution of $\delta_\mathrm{m} \approx 5.5\,\mu \mathrm{m}$.\\
The transmitted power is smaller by a factor of $0.18$ as compared to the theoretical expectations.
A possible reason for this deviation is an imperfect centering of the concave mirror with respect to the  fiber core of the outcoupling fiber.
The observed reduction would correspond to a small radial misalignment of $x_0 \approx 3.2\,\mu \mathrm{m}$ (see Section~\ref{sec:radalign}).
As can be seen in Fig. \ref{fig:fc_sym_asym} for cavity lengths $L>25\, \mu \mathrm{m}$ the observed reflectivity starts to drop and deviates from the predictions of the model introduced above.
This behaviour can be attributed to diffraction effects caused by the finite size of the concave mirror on the fiber tip. 
We model them by an effective reflectivity $R_2^\prime = R_2(1-l)$ of the outcoupling mirror, incorporating the clipping losses $l$ at the finite sized mirror with diameter $D_\mathrm{ref}=12.8\,\mu \mathrm{m}$ \cite{Hunger2010b} for $\sigma_\mathrm{ref}$.


As seen in Fig.\ \ref{fig:fc_sym_asym}a for $L~\approx~45\,\mu \mathrm{m}$ we observe a vanishing reflectivity on resonance $\sigma_\mathrm{ref} = 0$, which looks like an effectively symmetric cavity caused by losses ($R_2^\prime = R_1$). 
Increasing $l$ further reverses the asymmetry, leading to increasing $\sigma_\mathrm{ref}$ as captured by the model.\\

The measured transmission signal also deviates from the prediction of the analytical model for $L > 25\,\mu \mathrm{m}$. 
The faster decrease of the transmission compared to the model  can be explained by the smaller intra-cavity field caused by the clipping losses.

\subsection{Radial Alignment}
\label{sec:radalign}

\begin{figure}[t]
  \centering
    \includegraphics{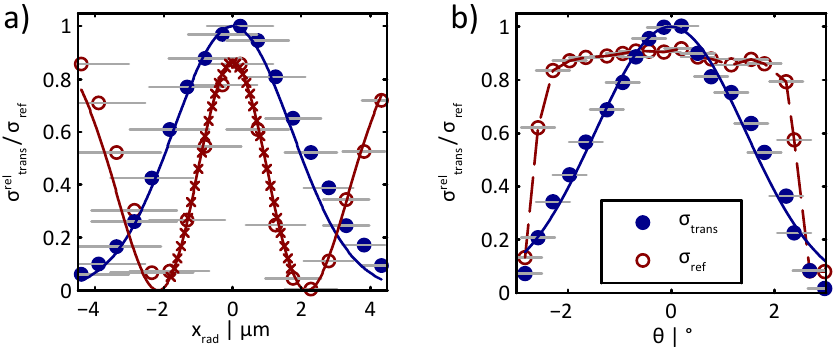}
     \caption{\label{fig:theta_x}\textbf{Sensitivity of an asymmetric fiber cavity with respect to misalignment}.
     \textbf{a)} Transmission and reflection signal for $L \approx 25\,\mu \mathrm{m}$ as a function Radial displacement. 
     The error bars correspond to the error of the micrometer drive calibration. 
     The solid lines are the expected behavior according to Eq.\ \ref{eq:betapcoffset}.
     The transmission data and model have been normailzed.
     Additionally, the crosses indicate a numerical simulation of the reflection signal with OSCAR.
     \textbf{b)} Angular dependence of the relative transmission and absolute reflection signal.
      The error bars correspond to the error of the angle determination caused by the limited resolution of the optical microscope  $\delta_\mathrm{m}$. 
     The solid blue line is a gaussian fit to the transmission signal.
     The transmission data and model have been normailzed.
     The dashed red line is a guide to the eye.}
\end{figure}

A radial misalignment  of the two fibers drastically changes the mode match as further discussed in App.~\ref{sec:gau_beam_overlap}. 
We experimentally measure the dependence of the mode match on radial misalignment as shown in Fig.\ \ref{fig:theta_x}a.
The relative position of the two fibers is independently determined with the microscope camera using  the fiber diameter as a length reference as above. 
A clear maximum at optimal alignment is visible. 
An important result is that the transmission signal with a FWHM of $ \Delta x^\mathrm{t}_\mathrm{rad} \approx 4.5\,\mu \mathrm{m}$ is much less sensitive to radial misalignment than the reflectivity on resonance with a FWHM of $ \Delta x^\mathrm{r}_\mathrm{rad} \approx 2\,\mu \mathrm{m}$. 
Our measurements show that it is important to align the asymmetric fiber cavity with sub-micrometer precision.
The model indicates that a radial misalignment of  $\pm \ 200\,\mathrm{nm}$ already leads to losses of 3.2 \%.\\
The decrease and subsequent increase of the reflectivity on resonance can be understood recalling the signal shown in Fig.\ \ref{fig:fc_sigma}a.
If the fiber is radially displaced, the mode match decreases and $\sigma_\mathrm{ref}$ drops.
Due to the minimum, $\sigma_\mathrm{ref}$ increases again if the radial displacement becomes large enough. 
Most of the light is promptly reflected off the incoupling mirror and is not interacting with the cavity resulting in a high $\sigma_\mathrm{ref}$.\\

\subsection{Angular Alignment}
We have also experimentally studied the robustness against angular misalignment. 
In this measurement the angle of the two fibers is changed by a defined value using a pitch/yaw stage and the cavity is realigned with a $xy$-translation stage.
We position the fiber end at the rotation point of the pitch/yaw stage allowing for a  measurement without changes of the cavity length. 
We calibrate the angles using a microscope camera as described above. 
The results are presented in Fig.\ \ref{fig:theta_x}b.
A large plateau ranging from $\pm 2^\circ$ for the reflectivity on resonance is visible. 
In contrast to the radial misalignment, the transmission with a FWHM of $ \Delta x^\mathrm{t}_\theta \approx 3.4^\circ$ is more sensitive to angular misalignment.
As a result a sub-degree angular alignment is necessary to ensure efficient transmission.\\
\section{Conclusion}
We have presented a thorough theoretical study and experimental realization of asymmetric fiber cavities.
The asymmetry poses new challenges with respect to the design of the cavity demanding for perfect mode match.
We have shown that planar concave cavities are ideally suited to realize such asymmetric cavities and to obtain high reflectivity on resonance,
ideally suited for e.g. use in mutually coupled quantum hybrid systems.
The mode field diameter of the light emitted from the incoupling fiber is the key parameter in designing the cavity and determines length $L$ as well as radius of curvature of the curved mirror $r_2$ within tight boundaries.
The requirements for excellent mode match also demand for very precise alignment due to the small mode field diameters involved.
The presented results show that radial position accuracy of $\pm 200\,\mathrm{nm}$ and a sub-degree angular alignment is necessary to ensure a minimal amount of loss,
which is also very favorable for usage in a cryogenic environment \cite{Zhong2014a}.

\subsection*{Acknowledgments} The authors acknowledge funding by the University of Hamburg and the German Research Foundation DFG under Grant No. BE 4793/2-1 and SE 717/9-1.

\appendix

\section{Overlap of Gaussian Beams}
\label{sec:gau_beam_overlap}
\subsection{Incoupling Mode Match $\mathbf{\beta}$}
The mode overlap $K_\mathrm{ax}(w_a,w_b,R_a,R_b)$ of two axially aligned gaussian beams can be calculated \cite{Goldsmith1998} when knowing the spot size $w$ and the curvature $R$ of the beams at a specific reference plane:
\begin{eqnarray}
	K_\mathrm{ax} &=& 4\left((w_b / w_a + w_a / w_b)^2 + \dots \right.\\
	& \dots& + \left. (\pi w_a w_b / \lambda)^2 (1/R_b - 1/R_a)^2 \right)^{-1}\label{eq:kax}.
\end{eqnarray}

If the waist of two beams is at the same position, which is used as the reference plane, the wave front curvature is $R = \infty$ for both beams. The overlap simplifies to
$$K_\mathrm{ax}(w_a,w_b) = \frac{4}{(w_b / w_a + w_a / w_b)^2}$$
for this case.
The waist $w_\mathrm{f}$ of the beam emitted by a planar cut fiber fiber is fixed by the $N\!A$ and the wavelength $\lambda$ used:
\begin{equation}
w_\mathrm{f} = \frac{\lambda}{0.82 \cdot \pi \cdot N\!A} \label{eq:wf}.
\end{equation}
For the incoupling efficiency $\beta^\mathrm{PC} = \beta^\mathrm{PC}\left(L,r_2\right)$ of a planar concave fiber cavity $w_a = w_\mathrm{f}$ whereas $w_b = w_0(L,r_2)$ is the cavity waist. 

\subsection{Transmission Mode Match $\mathbf{\eta_\mathrm{trans}}$}
To calculate the transmission mode match $\eta_2$ we have to consider the lensing effect of the outcoupling fiber, which is curved with $r_2$. It acts as a divergent lens and can be modeled as a thin lens with negative focal length $f = r_2 / \left( n_f - 1 \right)$ with $n_f$ being the refractive index of the fiber.

The curved surface changes only the wavefront curvature, not the spot size. The wavefront radius of curvature $R$ after refraction from the curved surface is therefore:
$R = \left[ \Re \left( 1/q_2 \right)  \right] ^{-1} = r_2/\left(n_f - 1\right).$
Now, $\eta_2$ can be calculated when taking the second mirror as the reference plane: 
\begin{equation}
	\eta_2 = K_\mathrm{ax}(w_2,w_\mathrm{f},r_2,-r_2/(n_f - 1)). \label{eq:eta2}
\end{equation}
Here, $w_2$ is the spot size of the cavity mode on the outcoupling mirror. The minus sign in Eq.\ \ref{eq:eta2} is a result of the orientation of the surface $r_2$.

\subsection{Radial Offset}
If a radial offset of length $x_0$ is introduced, the mode match is reduced by an additional exponential factor. In this case, the mode match $K_\mathrm{offset}$ \cite{Goldsmith1998} is
\begin{eqnarray}
K_\mathrm{offset} &=& K_\mathrm{ax} \exp \left[ -2 \left(\frac{x_0}{\delta_\mathrm{off}}\right)^2  \right], \mathrm{ with}\\
\delta_\mathrm{off} &=& \sqrt{ \frac{ \left( w_{0a}^2 + w_{0b}^2 \right)^2 + \left( \lambda \Delta z / \pi \right)^2}{ w_{0a}^2 + w_{0b}^2} } \label{eq:betapcoffset}
\end{eqnarray}
Here, $\delta_\mathrm{off}$ is the offset parameter and depends on the waists $w_{0a}$ and $w_{0b}$ and the axial distance between the location of the two waists $\Delta z$. For the case of a PC cavity, the waist of the fiber $w_\mathrm{f}$ and the waist of the cavity mode $w_0$ are at the same axial position, therefore $\Delta z = 0$.

The exponential reduction of the mode match with radial misalignment poses restrictions to the radial alignment precision necessary for the cavity.

\section{Numerical Simulations with OSCAR}
\label{sec:oscar}
\begin{figure}[t]
  \centering
    \includegraphics{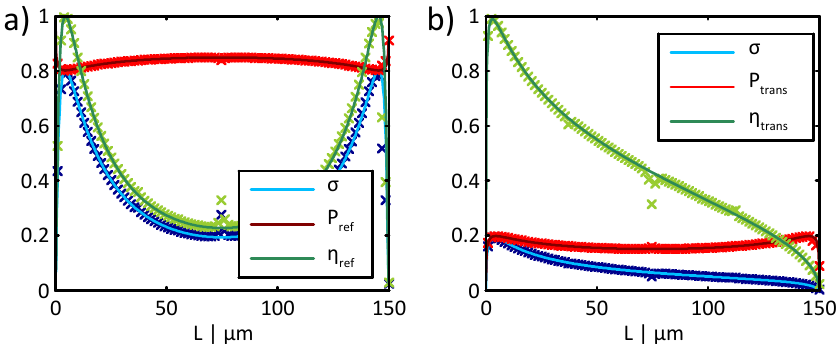}
     \caption{\label{fig:oscaranalytic} Comparison of the analytic solution (solid lines) for the observed signal $\sigma$, the reflected power $P_\mathrm{ref}$ and the mode match $\eta$ with the numerical simulation (crosses). Shown in \textbf{a)} is the reflection signal, \textbf{b)} the transmission signal for different cavity lengths.}
\end{figure}
To validate the model presented in the section above we also performed numerical simulations with the MATLAB script OSCAR\cite{Note4}. It is capable of calculating the spatial modes of cavities. We then calculate the overlap with the fiber mode to determine the amount of light transmitted in the fiber.\\
A comparison of the analytic expressions presented above for both the reflected and transmitted signal of a PC cavity with unstable mode match for different cavity lengths is shown in Fig.\ \ref{fig:oscaranalytic}. They show almost perfect agreement except at few distinct lengths ($L = r_2/2$, $L = r_2/4$, and $L = 3r_2/4$).\\

To gain further insight into the interference effect described in Eq.\ \ref{eq:spatialmode}, Fig.\ \ref{fig:cavityfields} shows the cavity field as calculated by OSCAR for the case of good and bad mode match. 
The circulating mode, located at the incoupling mirror, is larger than the incoupling mode, indicating the bad mode match. 
In contrast, the reflected mode, also shown at the incoupling mirror, develops a significant dip in the center due to the interference effect. 
As described above, this reflected mode has to be overlapped with the incoupling mode which results in a large losses.



%
%

%


\bibliography{library_submit}

\end{document}